# Dielectric Screening in Atomically Thin Boron Nitride Nanosheets


*Lu Hua Li,[1]\* Elton J. G. Santos,[2] Tan Xing,[1] Emmanuele Cappelluti,[3] Rafael Roldán,[4] Ying*

*Chen,[1]\* Kenji Watanabe[5] and Takashi Taniguchi[5]*

[1]Institute for Frontier Materials, Deakin University, Geelong Waurn Ponds Campus, Waurn Ponds, Victoria 3216, Australia

[2]Department of Chemical Engineering, Stanford University, Stanford, California 94305, United States

[3]Istituto dei Sistemi Complessi, CNR, v. dei Taurini 19, Rome 00185, Italy

[4]Instituto de Ciencia de Materiales de Madrid, CSIC, Sor Juana Ines de la Cruz 3, Madrid 28049, Spain

[5]National Institute for Materials Science, Namiki 1-1, Tsukuba, Ibaraki 305-0044, Japan


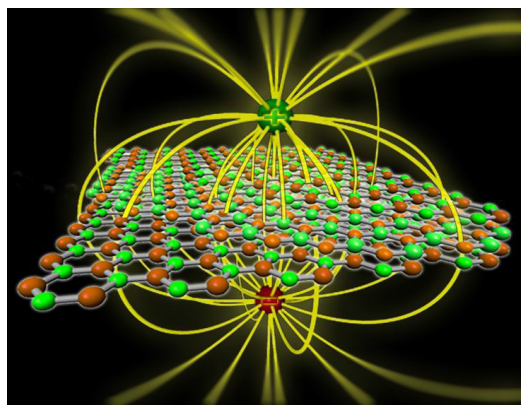





**ABSTRACT** Two-dimensional (2D) hexagonal boron nitride (BN) nanosheets are excellent dielectric substrate for graphene, molybdenum disulfide and many other 2D nanomaterials based electronic and photonic devices. To optimize the performance of these 2D devices, it is essential to understand the dielectric screening properties of BN nanosheets as a function of the thickness. Here, electric force microscopy along with theoretical calculations based on both state-of-the-art first-principles calculations with van der Waals interactions under consideration and non-linear Thomas-Fermi theory models are used to investigate the dielectric screening in high-quality BN nanosheets of different thicknesses. It is found that atomically thin BN nanosheets are less effective in electric field screening, but the screening capability of BN shows a relatively weak dependence on the layer thickness.

**KEYWORDS** boron nitride nanosheets, electric field screening, electric force microscopy (EFM), non-linear Thomas-Fermi theory, first-principles calculations



Two-dimensional (2D) nanomaterials, such as graphene, molybdenum disulfide ($MoS_2$) and tungsten diselenide ($WSe_2$) nanosheets, are promising candidates for next generation electronic and photonic devices of high-efficiency and flexibility.[1-6] The performance of these 2D devices highly relies on the dielectric substrate used. For example, the carrier mobility of graphene is far from satisfactory if it is placed on traditional dielectric substrate, such as silicon oxide ($SiO_2$), due to the scattering caused by the large roughness and trapped charges on the surface. Hexagonal boron nitride (BN) nanosheets are an excellent substrate to support graphene, $MoS_2$





and many other 2D nanosheets and can greatly improve their performance, due to its atomic flatness and low content of trapped charges.[3,6-8] In addition, laterally heterostructured electronics could be built by sandwiching graphene and other 2D materials between BN nanosheets.[3,9] Electric field screening is a fundamental and important property for any dielectric material. This is also true in the case of BN, because its dielectric screening could play a vital role in many applications, such as transport properties and control of gate voltage in graphene electronics.[10] Furthermore, the dielectric properties of underlying BN nanosheets can alter the dielectric constant of the graphene on top.[11] So it is highly desirable to have a deep understanding of the dielectric properties of BN nanosheets of different thicknesses for achieving 2D devices of optimal efficiency and performance with a miniaturized thickness. The dielectric properties of 15-19 nm thick polycrystalline BN nanosheets[12] and BN/graphene heterostructures,[13] all grown by chemical vapor deposition (CVD), have been measured recently. However, it is still unknown how the electric screening in BN nanosheets changes with the decrease of thickness. In this work, dielectric screening in atomically thin BN nanosheets of different thicknesses is analyzed using electric force microscopy (EFM) and theoretical calculations based on non-linear Thomas-Fermi theory models and first-principles calculations including van der Waals interactions. It is found that atomically thin BN nanosheets are less effective in electric field screening, but the screening capability of BN shows a relatively weak dependence on the layer thickness.

The BN nanosheets were mechanically exfoliated from hBN single crystals[14,15] and therefore of high-purity and high-quality.[16] An optical microscopy photo of the BN nanosheets of different thicknesses is shown in Figure 1a. Their layer numbers determined by an atomic force microscope (AFM) are in the range of 1-24 layers (L) (Figure 1b). The height trace of the dashed line in Figure 1b is shown in Figure 1c, where the 1L BN nanosheet has a thickness of 0.43 nm.





The measured thicknesses of the 2L and 24L nanosheets are 0.76 and 8.08 nm, respectively. Compared to the bulk hBN crystals, the atomically thin BN nanosheets of different thicknesses show different Raman shifts.[16,17] The Raman G band, corresponding to $E_{2g}$ vibration mode in hBN, of the 1L, 2L and 24L nanosheets are 1369.4, 1368.5 and 1366.9 $cm^{-1}$, respectively; while the G band of a bulk hBN crystal is 1366.4 $cm^{-1}$. These results are in line with our previous study.[16] The upshifts of the G band are due to the higher levels of strain and lower interlayer interactions in the atomically thin BN.

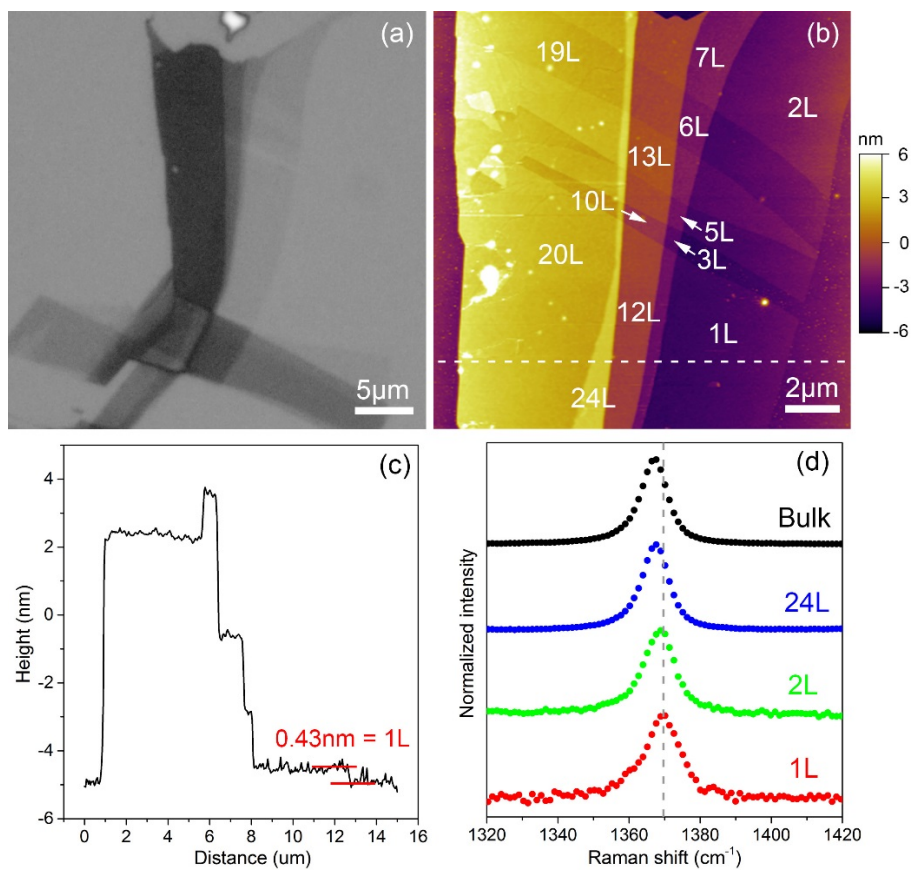

Figure 1. (a) Optical microscopy photo of BN nanosheets of different thicknesses on $SiO_2$/Si substrate; (b) AFM image of the BN nanosheets with the corresponding layer numbers labelled; (c) AFM height trace from the dashed line in AFM image in (b), showing that the thickness of





the 1L BN nanosheet is 0.43 nm; (d) Raman spectra of the 1L, 2L, 24L BN nanosheets and a bulk hBN crystal, where the dashed line shows the Raman shift of the 1L BN.

EFM was used to investigate the dielectric screening in the BN nanosheets. This method has been successfully used to study the electric field screening in graphene and $MoS_2$ nanosheets.[18,19] The EFM measurements were conducted in a lift mode, in which a conductive cantilever oscillating at its resonant frequency first scanned over the BN nanosheet for topography and then raised up at a constant height above the sample surface to detect long-range electric (*i.e.* capacitive) interactions between the cantilever and the surface of the BN nanosheets by monitoring the phase change of the cantilever. During the EFM measurements, various DC voltages were applied to the conductive cantilever tip ($V_{tip}$ in the range of +6 to –6 V).

The EFM images of the BN nanosheets in Figure 1b are shown in Figure 2. It can be seen that under a positive tip voltage ($V_{tip}$ = +6 V), the BN nanosheets show positive phase shifts (note that positive phase shift is defined as "attractive" in the Cypher AFM used in this work, which is opposite to the setting in AFMs from many other manufactures) relative to the $SiO_2$/Si substrate (Figure 2a); while under a negative voltage ($V_{tip}$ = –6 V), the BN nanosheets have relative negative phase shifts (Figure 2b). This peculiar phenomenon can be attributed to the water molecules confined between the BN nanosheets and $SiO_2$/Si substrate. The confined water molecules which come from air moisture adsorbed on hydrophilic $SiO_2$ have a preferable orientation to self-assemble into a dipolar film regardless of direction of external electric fields.[20] The dipolar water film serves as trapped charges in our EFM measurements, *i.e.* source of the charge doping in the BN nanosheets. EFM can measure the unscreened electric field from the





charge impurities in BN and therefore be used to investigate the dielectric screening properties of the BN nanosheets of different thicknesses, given that the intrinsic surface properties and trapped charges on the top surfaces of the BN nanosheets of various thicknesses are identical.

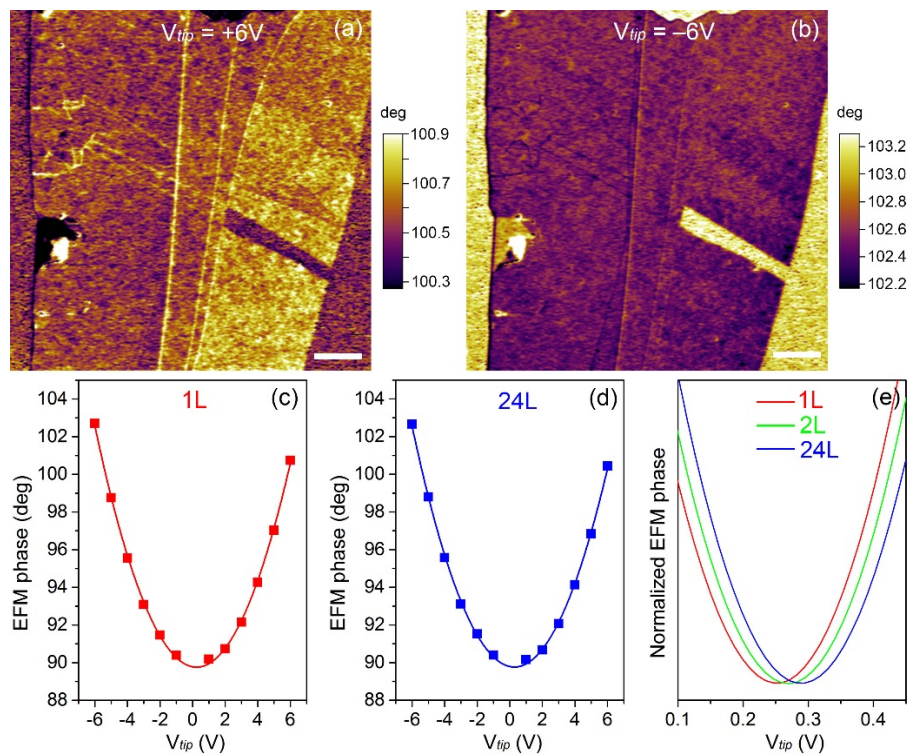

Figure 2. EFM phase images from the BN nanosheets of 1-24L under different $V_{tip}$: (a) +6 V and (b) −6 V; second-degree polynomial fittings to the EFM phase values of (c) 1L, (d) 24L BN under $V_{tip}$ ranging from +6 to −6 V; (e) comparison of the fitted phase shifts among 1L, 2L and 24L BN under $V_{tip}$ from 0.1 to 0.45 V.

The images under $V_{tip}$ = +6 and −6 V (Figure 2a and 2b) show that the EFM phase shifts are quite different for the 1L BN nanosheets and only slightly different among the 2-24L nanosheets. As EFM phase shifts directly relate to the unscreened charges from the confined water molecules





by the BN nanosheets, the observed differences in the EFM phase shifts indicate different capabilities of BN nanosheets of different thicknesses in electric field screening. Their differences in screening can be viewed more quantitatively by plotting the EFM phase shifts under different cantilever tip voltages. In our experiment, the cantilever frequency shift is smaller than the resonant frequency. So the EFM phase shift can be described as:[21]

$$\Delta\phi = \frac{\partial F/\partial z}{k} \cdot Q_{cant} \quad (1)$$

Where $\partial F/\partial z$ is the local force gradient or derivative of the electric force felt by the cantilever tip at a lift height $z$; $k$ is the spring constant of the cantilever; and $Q_{cant}$ is the Q factor of the cantilever. If the capacitive interaction between the conductive cantilever tip and the BN nanosheet is treated as an ideal capacitance, the force gradient of the capacitive interaction can be simply written as:[21]

$$\partial F/\partial Z = \frac{1}{2}\frac{\partial^2 C}{\partial z^2}\left(V_{tip} - V_s\right)^2 \quad (2)$$

Where $C$ is the local capacitance between the cantilever tip and the sample; $V_{tip}$ is the DC voltage applied to the cantilever tip; $V_s$ is the effective surface potential proportional to the unscreened charges on the BN surface. It can be seen from Eq. 2 that the EFM phase has a quadratic function to $V_{tip}$. So the measured EFM phase shifts of the BN nanosheets under $V_{tip}$ ranging from −6 to +6 V can be well fitted using second-degree polynomials (Figure 2c and 2d). The comparison of the fitted EFM phase shifts from the BN nanosheets of different thicknesses clearly show that their force gradients ($\partial F/\partial z$) of the capacitive interactions are different and the vertexes of the fitted parabolas (*i.e.* phase shift minima or $V_s$) are shifted to higher values with the increased thickness





of the BN nanosheets (Figure 2e), representing the different dielectric screening in the BN nanosheets of different thicknesses.

The deviations of the $V_s$ values for the BN nanosheets of different thicknesses from the bulk hBN crystal (*i.e.* $\Delta V_s = V_{s(bulk)} - V_s$) are summarized in Figure 3a (black dots). It can be seen that the screening capability of BN decreases with the reduction of its thickness. However, compared to those in graphene and $MoS_2$ nanosheets,[18,19] such decrease in screening in BN is much less steep. In other words, the dielectric screening in atomically thin BN nanosheets has a relative weak dependence on their thickness. Nevertheless, the screening of monolayer BN is much weaker than the thicker BN nanosheets (Figure 3a). In fact, the difference in $\Delta V_s$ between monolayer and bilayer BN roughly equals to the difference in $\Delta V_s$ between bilayer and 24L BN.

The screening behavior in BN nanosheets can be described by nonlinear Thomas-Fermi theory which has been used to analyze the electric screening in graphene and $MoS_2$ nanosheets.[18,19] The surface charge density of the underlying water dipolar film can be defined to be $-\sigma_0$ and the BN has a charge density of $\sigma_0$. The charge distribution $\sigma_{(z)}$ inside BN can therefore be derived by minimizing the grand potential of the system via the balance between the kinetic energy of charge carriers and electrostatic potential between BN layers. Then, $\sigma_{(z)}$ can be used to deduce a dimensionless screening parameter, $r_D$ and ultimately the surface potential difference between the top and bottom of BN, $\Delta V(D)$.

The scenario when interlayer coupling or hopping is excluded is firstly considered. BN has a parabolic conduction band. So in this 2D model, the difference in surface potential between the top and bottom layer of the BN can be described as:[19]





$$\Delta V(D) = \frac{2\pi \hbar^2 \sigma_0}{e N_s N_v m_\parallel} \sqrt{2\beta_0} d \; \frac{1 - r_D}{\sqrt{1 - r_D^2}} \quad (3)$$

Where $\hbar$ is the reduced Plank constant; $N_s$ and $N_v$ are the spin and valley degeneracies ($N_s = N_v = 2$); $m_\parallel$ is the in-plane effective mass; $d$ is the interlayer spacing in hBN ($d = 0.334$ nm); $\beta_0 = e^2 N_s N_v m_\parallel / 4\pi\varepsilon_0 \kappa_\perp \hbar^2$, and $\varepsilon_0$ and $\kappa_\perp$ are the permittivity of free space and the dielectric constant of hBN in the $c$-direction (we set $\kappa_\perp = 4$),[12] respectively. $r_D$ can be solved from the nonlinear equation:

$$r_D = \frac{\sigma_{(D)}}{\sigma_{(0)}} = \frac{1}{\cosh(\sqrt{\frac{2\beta_0}{d}} D)} \quad (4)$$

Where $\sigma_{(0)}$ and $\sigma_{(D)}$ are the charge densities at the bottom and the top of a BN nanosheet with a thickness $D$. Note that the Thomas-Fermi theory is valid only when $D$ is much larger than $d$ ($D \gg d$). The experimental results from EFM can be fitted with $\sigma_0$ and $m_\parallel$ (which is confined to $0.01 m_e$) as two variables. The blue dashed line in Figure 3a shows the result from the 2D fitting when the values for $\sigma_0$ and $m_\parallel$ are $2\times10^{12}$ cm$^{-2}$ and $0.01 m_e$. According to the fitting, BN nanosheets should be mostly in a weak coupling regime (except when D < 1 nm) where the kinetic energy surpasses the electrostatic energy, as $\sqrt{2\beta_0}/_d \, D > 1$. In other words, the 2D fitting predicts that the screening in BN should drop sharply with the decrease of thickness. This is starkly different to the experimental observation that BN of reduced thicknesses show a very gradual decrease in screening. Hence, interlayer coupling or hopping should play an important role in the screening in BN. When interlayer coupling is considered, a 3D model can be established:[19]



$$\Delta V(D) = \frac{1}{2}\left(\frac{6\pi^2\hbar^3}{N_s N_v dm_\parallel\sqrt{m_\perp}}\right)^{2/3}\left(\frac{25\beta_\perp d\sigma_0{}^2}{8e^2}\right)^{2/5}\frac{1-r_D}{\left(1-r_D{}^{5/2}\right)^{2/5}} \quad (5)$$

Where $N_s = 2$ and $N_v = 3$ in normal AA' BN stacking; $m_\perp = \hbar^2/2|t_\perp|d^2$ and $t_\perp$ is an interlayer hopping parameter, whose absolute values are in the range of 0.25-0.32 eV;[22,23] $\beta_\perp = (4e^2/5\varepsilon_0\kappa_\perp)$ $(N_s N_v dm_\parallel\sqrt{m_\perp} \,/\, 6\pi^2\hbar^3)^{2/3}$; and $r_D$ can be numerically determined from:

$$\left[\frac{25\beta_\perp d\sigma_0{}^2}{8\left(1-r_D{}^{5/2}\right)}\right]^{-1/10}\int_{r_D}^1\frac{du}{\left(u^{5/2}-r_D{}^{5/2}\right)^{1/2}} = \sqrt{\frac{2\beta_\perp}{d}}D \quad (6)$$

The red solid line in Figure 3a shows the 3D fitting in which $\sigma_0$ and $m_\parallel$ are $1.3\times10^8$ cm$^{-2}$ and $0.1m_e$, respectively (see Supporting Information). The 3D model which has a much better consistency to the EFM results indicates that BN is mostly in a strong coupling regime. This can be better seen in Figure 3b. The distributions of charge density as a function of thickness [$\sigma_{(z)}$] inside 10, 20 and 40 nm thick BN could also be estimated according to the 3D model (Figure 3c).

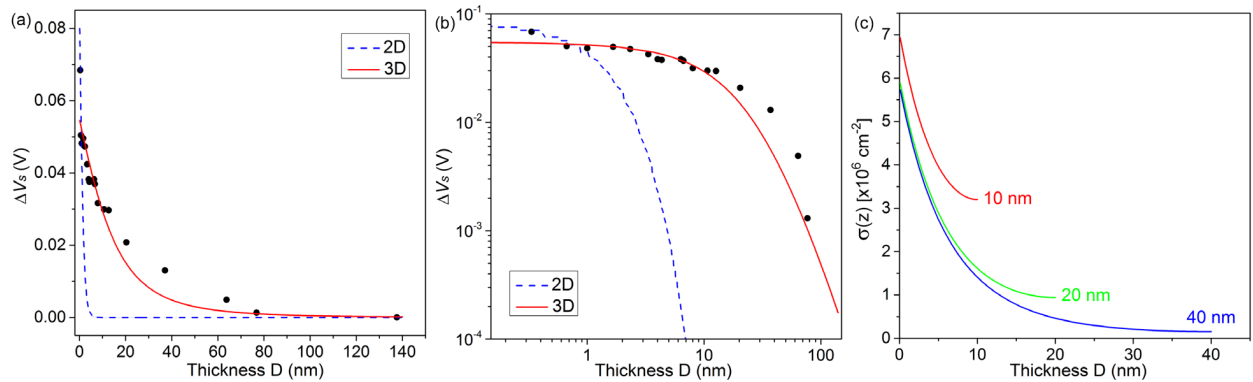

Figure 3. (a) The EFM derived deviations of the $V_s$ values ($\Delta V_s$) of BN nanosheets of different thicknesses from the bulk hBN crystal caused by the unscreened charges as well as fittings using 2D (dash in blue) and 3D (solid in red) non-linear Thomas-Fermi theory; (b) similar to (a) but on





logarithmic scales; (c) predicted charge distributions σ(z) inside BN nanosheets of different

thicknesses, *i.e.* 10, 20 and 40 nm, according to the 3D non-linear Thomas-Fermi theory.

The continuum limit of Thomas-Fermi theory makes it less suitable for studying electric field

screening in extremely thin BN. Therefore, first-principles total energy calculations were

performed on atomically thin BN of different thicknesses under an external electric field using

density functional theory, which can simulate the applied gate voltage at the BN surfaces. The

application of an external gate bias induces an interlayer charge transfer that partially screens the

external field, generating effective electric field ($E_{eff}$) between the layers. Figure 4 shows the

distribution of $E_{eff}$ as a function of the distance $z$ under an external field of 0.0105 V/Å across the

supercell. We focus on the electrical response of 1L, 2L, 3L and 6L BN nanosheets, which

captures the essential features of their discrete screening properties. It can be seen that $E_{eff}$

depends on position with its maximum value at the midpoint between the layers. Relative

oscillations of the minimum values of $E_{eff}$ are observed from thicker BN, namely 3L and 6L

(Figure 4). This can be attributed to slight variations in the interlayer distance in multilayered BN

whose binding energy between the layers is reduced under a finite field.[24,25] It should be also

noted that $E_{eff}$ does not change appreciably as the layer thickness increases from 1L to 6L.

Calculations at thicker BN structures (up to 54L) show a similar behavior. The calculated

dielectric constants of 1L, 2L and 3L BN are 2.31, 2.43 and 2.49, respectively. This indicates

that atomically thin BN of different thicknesses has quite similar capability in electric field

screening, in good agreement with the EFM results which also show that thickness has a weak

effect on the measured unscreened charges on few-layer BN nanosheets, except the 1L one. The

much weaker screening in the 1L BN observed from the EFM measurements is probably due to





much more charge transfer between the underlying dipole water film and the 1L BN caused by electrons in the π orbitals located above and below a single BN plane. So, the dielectric screening in BN is very different to that in graphene which shows a quick decay of effective electric field in the first few layers and somewhat similar to that in $MoS_2$.[24,25]

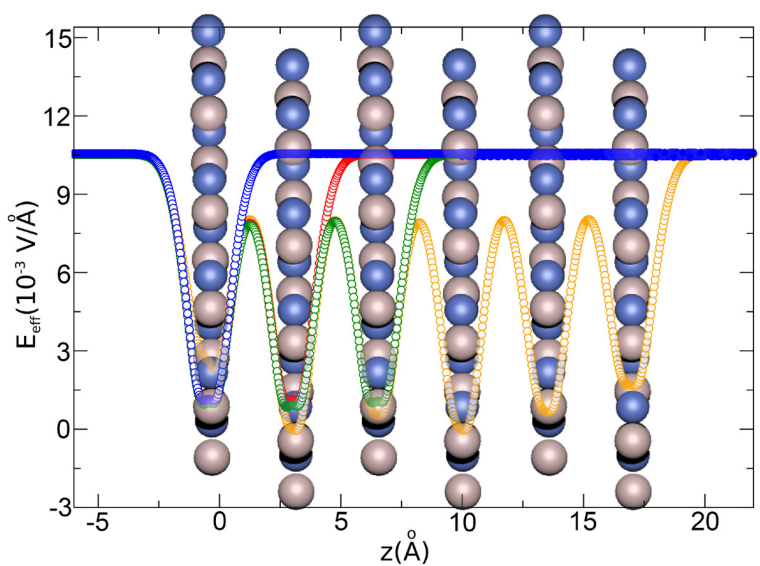

Figure 4. First principles calculated effective electric fields ($E_{eff}$) inside 1L (blue), 2L (red), 3L (green) and 6L (orange) BN nanosheets under an external field of 0.0105 V/Å. The balls represent boron and nitrogen atoms in the BN layers.

In summary, the electric field screening in BN of different thicknesses has been studied by EFM and theoretical calculations. It is found that the screening in atomically thin BN shows a weak dependence on its thickness, which is in line with the smooth decay of electric field inside few-layer BN revealed by the first-principles calculations. According to the non-linear Thomas-Fermi theory, BN is mostly in a strong coupling regime and interlayer hopping plays a crucial role in the screening. The electric field screening in BN nanosheets of different thicknesses





presented here make a step forward in designing and optimizing 2D electronic and optoelectronic devices where BN nanosheets are used as dielectric substrate to support graphene, $MoS_2$ and other 2D nanomaterials.

**Methods.** The atomically thin BN nanosheets were produced by Scotch tape exfoliation of single crystal hBN[14] on 90 nm $SiO_2$/Si substrate and visualized using an Olympus BX51 optical microscope equipped with a DP71 camera. A Cypher scanning probe microscope (Asylum Research) was used for the AFM and EFM measurements using a Pt/Ti coated conductive cantilever with a spring constant of 2 N/m (AC240TM, Asylum Research). The lift height for all EFM measurements was 30 nm. A Renishaw inVia confocal micro-Raman spectrometer equipped with a 514.5 nm laser (25 mW) was used to characterize the molecular vibrations in BN nanosheets of different thicknesses.

Calculations are based on *ab initio* density-functional-theory using the SIESTA code.[26] The generalized gradient approximation[27] and nonlocal van der Waals density functional[28] were used together with double-$\zeta$ plus polarized basis set, norm-conserving Troullier-Martins pseudopotentials[29] and a mesh cutoff of 150 Ry. Atomic coordinates were allowed to relax using a conjugate-gradient algorithm until all forces were smaller in magnitude than 0.01 eV/Å. Relevant lattice constants (in-plane and out-of-plane) were optimized for each system. To model the system studied in the experiments, we created supercells containing up to 108 atoms to simulate multilayer BN systems. To avoid interactions between supercell images the distance between periodic images of the BN structures along the direction perpendicular to the BN-plane was always larger than 20 Å. The resolution of the real-space grid used to calculate the Hartree





and exchange-correlation contribution to the total energy was chosen to be equivalent to 150 Ry plane-wave cutoff. The number of $k$-points was chosen according to the Monkhorst-Pack scheme,[30] and was set to the equivalent of a 45x45x1 grid in the two-atom primitive unit cell of BN, which gives well converged values for all the calculated properties. We used a Fermi-Dirac distribution with an electronic temperature of $k_B T$=21 meV. The external electric field is introduced through a saw-tooth-like electrostatic potential in the direction perpendicular to the BN plane.

ASSOCIATED CONTENT

**Supporting Information.** Charge density in BN nanosheets estimated by scanning Kevin probe microscopy. This material is available free of charge via the Internet at http://pubs.acs.org.

AUTHOR INFORMATION

**Corresponding Author**


*Email: luhua.li@deakin.edu.au; ian.chen@deakin.edu.au


**Author Contributions**

L.H.L. designed the research, produced the material, conducted EFM measurements and involved in data analysis and results discussion. E.J.G.S. did first-principles calculation. T.X., Y.C., E.C. and R.R. involved in data analysis and result discussion. K.W. and T.T. provided single crystal hBN. L.H.L. and E.J.G.S. co-wrote the manuscript with input from all the other authors.





**ACKNOWLEDGMENT**

L.H.L. thanks Alfred Deakin Postdoctoral Research Fellowship and Australian Research Council Discovery Program for the financial support. E.J.G.S. thanks the computational resources provided by the Extreme Science and Engineering Discovery Environment (XSEDE), supported by NSF grant numbers TG-DMR120049 and TG-PHY120021. E.C. acknowledges support from the European project FP7-PEOPLE-2013-CIG "LSIE_2D", Italian National MIUR Prin project 20105ZZTSE and the Italian MIUR program "Progetto Premiale 2012" Project ABNANOTECH. R.R. acknowledges financial support from the Juan de la Cierva Programe and Grant No. FIS2011-23713 (MINECO, Spain).